\documentstyle[aaspp4,12pt,flushrt]{article}
\def\kms{km~s$^{-1}$}
\def\mzw{m_{\rm Zw}}
\begin{document}
\title{A Complete Redshift Survey to the Zwicky Catalog Limit\\
in a $2^{\rm h} \times 15\arcdeg$ Region Around 3C 273}

\author{Norman A. Grogin, Margaret J. Geller, and John P. Huchra} 

\affil{Harvard-Smithsonian Center for Astrophysics,
60 Garden Street, Cambridge, MA 02138\\ 
E-mail: ngrogin, mgeller, jhuchra@cfa.harvard.edu} 

\slugcomment{Accepted for publication in the {\it Astrophysical Journal, Supplement Series}} 

\abstract{We compile 1113 redshifts (648 new measurements, 465 from
the literature) for Zwicky catalogue galaxies in the region ($-3\fdg5
\leq \delta \leq 8\fdg5$, $11\fh5 \leq \alpha \leq 13\fh5$).  We
include redshifts for 114 component objects in 78 Zwicky catalogue
multiplets.  The redshift survey in this region is 99.5\% complete to
the Zwicky catalogue limit, $\mzw=15.7$.  It is 99.9\% complete to
$\mzw = 15.5$, the CfA Redshift Survey (CfA2) magnitude limit.

The survey region is adjacent to the northern portion of CfA2,
overlaps the northernmost slice of the Las Campanas Redshift Survey,
includes the southern extent of the Virgo Cluster, and is roughly
centered on the QSO 3C 273.  As in other portions of the Zwicky
catalogue, bright and faint galaxies trace the same large-scale structure.  
}


\begin{section}{Introduction}
The Center for Astrophysics Redshift Survey (hereafter CfA2,
\cite{gh89}) of galaxies from Zwicky's Catalogue of Galaxies and
Clusters of Galaxies (\cite{cgcg}, hereafter CGCG) still remains one
of the best resources for the study of the nearby galaxy distribution.
In a recent paper (\cite{gg98}), we used CfA2 to investigate the
connection between local ($cz \lesssim 10000$~\kms) Lyman-alpha
absorption systems observed with {\sl HST} (\cite{bah91,mor91,wey95,ssp95,ssp96}) 
and the large-scale structure in the galaxy
distribution.  To explore the large-scale structure around the
Ly$\alpha$ absorbers in the foreground of 3C 273, we extended the
redshift survey of CGCG galaxies into a wide region around 3C 273
complete to the catalogue magnitude limit, $\mzw=15.7$

In contrast with earlier pencil-beam redshift surveys around
Ly$\alpha$ absorption systems (\cite{mor93,l95}, and others), we
surveyed a much shallower but more extended area, $2^{\rm
h}\times12\arcdeg$, around 3C 273.  There are 788 CGCG galaxies in the
survey region ($11\fh5 \leq \alpha_{1950} \leq 13\fh5$, $-3\fdg5 \leq
\delta_{1950} \leq 8\fdg5$) to $\mzw=15.5$, the CfA2 magnitude limit,
and 1082 with $\mzw\leq15.7$.  In \S\ref{obssec} we describe the
observations and the redshift catalogue included here.  We briefly
comment on the redshift distribution in \S\ref{discsec}.
\end{section}
\begin{section}{Observations and Redshift Catalogue} \label{obssec}
Among the 1082 CGCG entries in the survey region, ZCAT (\cite{zcat})
and NED\footnote{The NASA/IPAC Extragalactic Database (NED) is
operated by the Jet Propulsion Laboratory, California Institute of
Technology, under contract with the National Aeronautics and Space
Administration.} contain 465 previously measured redshifts that we
include in our catalogue.  We obtained spectra of the remainder with
the 1.5m Tillinghast telescope at the F. L. Whipple Observatory.  We
present 81 new redshifts from observations with the Z-machine
spectrograph (\cite{zmach}) prior to 1994.  For the Z-machine
observations, the exposures are typically 15--50 minutes with a 600
line mm$^{-1}$ grating, providing spectral coverage from
4500--7100\AA\ at 5\AA\ resolution.  We present 566 new redshifts from
observations with the FAST spectrograph (\cite{FAST}) since 1994.  The
FAST observations are typically 5--20 minute exposures with a 300 line
mm$^{-1}$ grating, providing spectral coverage from 3600--7600\AA\ at
6\AA\ resolution.  The resulting Reticon (Z-machine) and CCD (FAST)
spectra were re-reduced and wavelength-calibrated as part of the CfA
spectroscopic data pipeline (cf.~Kurtz \& Mink 1998, hereafter
\cite{km98}).

We determine radial velocities by absorption and emission template
cross-correlation, using the IRAF task XCSAO in the RVSAO package
(\cite{km98}).  We cross-correlate the Z-machine spectra with {\sl
ztemp}, a composite absorption-line spectrum (\cite{td79}), 
and with {\sl femtemp97}, a recently-developed synthetic
emission line template (\cite{km98}).  We cross-correlate the FAST
spectra with {\sl femtemp97} and with {\sl fabtemp97}, a composite
absorption line template introduced in \cite{km98}.  We estimate the
uncertainty in the {\sl femtemp97} and {\sl fabtemp97} velocities from
the respective error models in \cite{km98}:
\begin{eqnarray}
{\sl femtemp97}:\quad \Delta v  &=& \left[{20^2/2 + 
220^2/(1+r)^2}\right]^{1/2} {\rm km\ s}^{-1}, \nonumber\\
{\sl fabtemp97}:\quad \Delta v  &=& \left[{20^2/2 + 
360^2/(1+r)^2}\right]^{1/2} {\rm km\ s}^{-1},
\end{eqnarray}
where $r$ is the cross-correlation statistic (\cite{td79}).  
For the uncertainty in the {\sl ztemp} velocities, 
we follow the  prescription (\cite{falc98}): 
\begin{equation}
{\sl ztemp}:\quad \Delta v = \left[{35^2 + 
212^2/(1+r)^2}\right]^{1/2} {\rm km\ s}^{-1}.
\end{equation}
We choose the velocity of the template with the smallest $\Delta v$
unless the velocities for the absorption and emission templates differ
by $<300$ \kms.  In the case of a velocity difference $<300$ \kms, we
use the error-weighted mean of the absorption and emission
template velocities.  The velocities and their errors presented here
supersede previously published CfA measurements for the same galaxies.

Table \ref{cztab} lists all 1082 CGCG entries with $11\fh5 \leq
\alpha_{1950} \leq 13\fh5$ and $-3\fdg5 \leq \delta_{1950} \leq
+8\fdg5$.  We order the galaxies by CGCG coordinate designation and
provide the corresponding NGC or IC number, if available, in the
comment field.  We determine the galaxy coordinates to
$\sim\!\pm2\arcsec$ using the Digitized Sky Survey (hereafter DSS,
\cite{DSS}).  All tabulated coordinates are epoch J2000; all tabulated
redshifts are heliocentric.  We provide a reference number for the
source of each redshift and list the sources separately in Table
\ref{velsrc}.  The new Z-machine and FAST velocities also include the
following suffixes in their velocity source column: ``z'' to denote a
{\sl ztemp} match; ``a'' to denote a {\sl fabtemp97} match; and ``e''
to denote a {\sl femtemp97} match.  We include five galaxies which
have no previously published redshift and whose FAST spectra were too
weak for us to obtain a reliable ($r\gtrsim3$) cross-correlation match
with either template.  These galaxies appear in Table \ref{cztab} with
a position but no velocity.

In Table \ref{mult} we catalogue 78 CGCG entries in our survey region
which are close associations of fainter sources that we have split by eye
using the DSS.  Table \ref{mult}
includes all entries noted by Zwicky as compound systems except for two
(1158.6$-$0100 and 1223.7$-$0101) that we were unable to subdivide.  
We name the components with a suffix indicating the geometry of
the system: ``E'' and ``W'' for a primarily east-west pair, ``N'' and
``S'' for a north-south pair, or alphabetically in order of increasing
right ascension for a system with more than two components.  As in
Table \ref{cztab}, we provide arcsecond coordinates from the DSS and
heliocentric velocities with reference codes enumerated in Table
\ref{velsrc}.  

We have a redshift for the brightest member of every multiplet in
Table \ref{mult}, and present a total of 114 multiplet redshifts (85
new, 29 from the literature).  One of us (JPH) has estimated component
magnitudes for many of these multiplets by eye (cf.~ZCAT).  If one
member of a CGCG multiplet predominates in brightness, we also list
its coordinates and redshift in Table \ref{cztab} and note it as a
multiplet component in the comment field.  We otherwise leave blank
the coordinates and redshift in Table \ref{cztab}, only listing the
CGCG magnitude and ``multiple'' in the comment field.  Because Zwicky
assigned one magnitude to each multiplet based on the combined light,
it is not surprising that the component redshifts in Table \ref{mult}
are generally larger than the redshifts for non-compound CGCG entries
of comparable magnitude.

The small velocity differences among components in Table \ref{mult}
suggest that most of the CGCG multiplets are physically associated and
not chance superpositions.  We select the 17 multiplets from Table
\ref{mult} with component velocity differences $<300$ \kms\ and
present them in Figure \ref{multfig} as a mosaic of $4\arcmin$ DSS
fields.  Several of these systems (e.g.~1334.0+0106 and 1224.5-0037)
are of interest as merger candidates, and three others (1131.5-0319,
1237.9+0827, and 1300.4+0547) may be
compact groups (cf.~\cite{hcg,rscg}).

If we count the multiplets as one object apiece, this survey is
currently 99.9\% complete (787 of 788) to the CfA2 magnitude limit,
$\mzw=15.5$.  The survey has redshifts for 99.5\% of the galaxies
(1077 of 1082) to the CGCG magnitude limit, $\mzw=15.7$.
\end{section}
\begin{section}{Discussion} \label{discsec}
Figures \ref{wedgelo} and \ref{wedgehi} show the redshift-space
distribution of CGCG galaxies from Table \ref{cztab} in ``slices'' of
$6\arcdeg$-declination thickness, centered on $\delta_{1950}=-0\fdg5$
and $\delta_{1950}=+5\fdg5$, respectively.  We plot the $\mzw\leq15.5$
galaxies as open squares, and the $\mzw>15.5$ galaxies as crosses.
Roughly 3\% of the galaxies have $cz>20000$ \kms\ and thus lie outside
the radial boundary of the plots.

Figure \ref{velhist} shows the histogram of radial velocities from
Table \ref{cztab}.  Because Zwicky only claimed his catalogue was
complete to $\mzw=15.5$, we plot the velocity histogram for
$\mzw\leq15.5$ galaxies (solid line) in addition to the histogram for
the entire $\mzw \leq 15.7$ sample (dashed line).  The peak at
$cz\sim1000$--2000 \kms\ is the southern extent of the Virgo Cluster,
appearing as a narrow radial ``finger'' in Figure \ref{wedgehi}.  The
sharp peak in galaxy counts at $cz\sim6000$--7000 \kms\ is the
continuation of the ``Great Wall'' first observed across the northern
CfA Redshift Survey (\cite{gh89}).  The enhancement in galaxy counts
is spread across the entire R.A.  range, persists across both
declination slices, and matches the redshift of the Great Wall at this
R.A. at larger declination.

In an earlier study of the faintest CGCG galaxies, Thorstensen et
al.~(1995) measured redshifts for 241 $\mzw>15.5$ galaxies in the
original CfA ``Slice of the Universe'' ($8^{\rm
h}\leq\alpha_{1950}\leq17^{\rm h}$, $26\fdg5\leq\delta_{1950}
\leq32\fdg5$) and found that the fainter galaxies outlined the same
structures seen in the $\mzw\leq15.5$ survey.  We similarly observe
that the 277 $\mzw>15.5$ galaxies in Figures \ref{wedgelo} and
\ref{wedgehi} generally trace the same structures as the 769
$\mzw\leq15.5$ galaxies.  For example, there is a large region in
Figure \ref{wedgelo} ($12^{\rm h}15^{\rm m} \lesssim \alpha_{1950}
\lesssim 13^{\rm h}15^{\rm m}$, $7500 \lesssim cz \lesssim 14000$
\kms) equally devoid of both brighter and fainter CGCG galaxies.
Interestingly, there are two Ly$\alpha$ absorption systems within this
prominent void in the 3C 273 foreground.  We discuss the implications
at length in Grogin \& Geller (1998).

\acknowledgements We acknowledge P. Berlind, G. Hazenberg, and
J. Peters for making many of the 1.5m observations; E. Falco,
M. Kurtz, J. Mader, D. Mink, S. Tokarz, and C. Wu for assistance with
the data reduction; and D. Fabricant for the FAST spectrograph.  We
also acknowledge suggestions from the scientific editor, G. Bothun,
regarding the CGCG multiplets.  This research was supported in part by the
Smithsonian Institution.
\end{section}
\newpage
\dummytable
\label{cztab}
\newpage
\dummytable
\label{velsrc}
\newpage
\dummytable
\label{mult}

\clearpage
\begin{center} {\large Figure Captions} \end{center}
\medskip
\figcaption[south.ps]{ Wedge diagram showing the 480 Zwicky catalogue
galaxies from Table \ref{cztab} with measured redshifts in the
southern half of the survey region, $-3\fdg5 \leq \delta_{1950} \leq
2\fdg5$.  Of these, 8 galaxies have $cz > 20000$ \kms\ and lie beyond
the plotted region.  Galaxies with $\mzw \leq 15.5$ are plotted as
open squares; fainter galaxies to the Zwicky catalogue limit ($\mzw =
15.7$) are plotted as crosses.
\label{wedgelo}
}

\figcaption[north.ps]{ Wedge diagram showing the 566 Zwicky catalogue
galaxies from Table \ref{cztab} with measured redshifts in the
northern half of the survey region, $2\fdg5\leq \delta_{1950} \leq
8\fdg5$.  Of these, 21 galaxies have $cz > 20000$ \kms\ and lie beyond
the plotted region.  Galaxies with $\mzw \leq 15.5$ are plotted as
open squares; fainter galaxies to the Zwicky catalogue limit ($\mzw =
15.7$) are plotted as crosses.
\label{wedgehi}
}

\figcaption[velhist.ps]{ Velocity histograms of CGCG galaxies from
Table \ref{cztab} in the survey region ($-3\fdg5 \leq \delta_{1950}
\leq 8\fdg5$, $11\fh5 \leq \alpha_{1950} \leq 13\fh5$): 1046 $\mzw \leq
15.7$ galaxies (dotted line) and the subset of 769 $\mzw \leq 15.5$
galaxies (solid line).
\label{velhist}
}

\figcaption[multfig.ps]{Mosaic of the 17 CGCG multiplet fields from
Table \ref{mult} with measured component velocities separated by
$<300$ \kms.  Images are $4\arcmin \times 4\arcmin$ and extracted from
the Digitized Sky Survey.  All images are oriented with North to the
top and East to the left.
\label{multfig}
}
\end{document}